\newcommand{\rem}[1]{}

\documentclass[prb,twocolumn,amsmath,amssymb]{revtex4}
\usepackage{amsmath,graphicx,subfigure}

\begin{document}
\title{Spectral Curves and Localization\\
in Random Non-Hermitian Tridiagonal Matrices}
\author{L.~G.~Molinari$^{(1,2)}$ and 
G.~Lacagnina$^{(2)}$\\
${}^1$ Universit\`a degli Studi di Milano, Dipartimento di Fisica,
Via Celoria 16, Milano I-20133;\\
${}^2$ I.N.F.N. Sezione di Milano, Via Celoria 16, Milano I-20133.}
\date{May 2009}
\begin{abstract}
Eigenvalues and eigenvectors of non-Hermitian tridiagonal periodic random 
matrices are studied by means of the Hatano-Nelson deformation. 
The deformed spectrum is annular-shaped, with inner radius measured by the 
complex Thouless formula. The inner bounding circle 
and the annular halo are stuctures that correspond to the two-arc and 
wings observed by Hatano and Nelson in deformed Hermitian models, and
are explained in terms of localization of eigenstates via a spectral duality 
and the Argument principle.
\end{abstract}
\maketitle

\section{Introduction}
Hermitian tridiagonal random matrices are studied in great detail, and
many results are available on spectral properties such as density, 
statistics and localization of eigenvectors. They appear in several models
of physics, as Dyson's random chains, Anderson's models for 
transport in disordered potentials, Ising spin models with random couplings, 
$\beta$-ensembles of tridiagonal random matrices. Hatano and 
Nelson\cite{Hatano96} introduced a beautiful method to study the 
localization of 
eigenvectors, by forcing an asymmetry of upper and lower nondiagonal
elements. Then the eigenvalues are driven from 
the real axis to curves in the complex plane, in patterns that
measure the localization length of the corresponding eigenvectors. 

Tridiagonal random matrices that are non-Hermitian from the start are less 
studied. They model systems with asymmetric hopping 
amplitudes\cite{Derrida00,Goldsheid00,Zee03}, describe properties of 1D
random walks\cite{Cicuta00,Bauer02} or the evolution of population
biology\cite{Shnerb98}. Their spectrum is complex. In this work we study 
how the Hatano-Nelson deformation modifies it, the occurrence of spectral 
curves and the connection with the localization of eigenvectors.

\begin{figure}
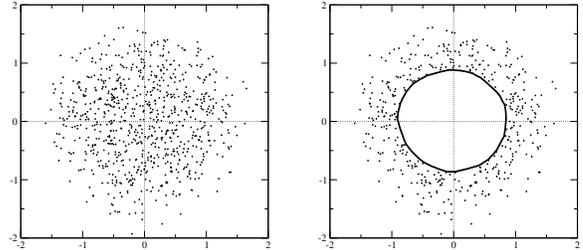

\begin{center}
\includegraphics[width=3.5cm]{plot_N800_z_1_0.eps}\hspace{0.5cm}
\includegraphics[width=3.5cm]{plot_N800_z_sqrtE_0.eps}
\caption{\label{a1} 
Eigenvalues in the complex plane of a single non Hermitian
tridiagonal matrix of size $n=800$. The same matrix entries $\{a_k,b_k,c_k\}$
are used, with $\xi =0$ and $\varphi=0$ (left), and $\xi=0.5$ and 
$\varphi =0$ (right).
Note that the eigenvalues out of the circle have the same positions
in the both cases.}
\end{center}
\end{figure}

Let us then consider complex tridiagonal matrices with corners
\begin{eqnarray}
M =\left [\begin{array}{cccc}
a_1 &    b_1   &   {}  & c_1\\
c_2 & \ddots &\ddots   & {}    \\
{}  & \ddots & \ddots  & b_{n-1}   \\
b_n&    {}  &   c_n   & a_n   \\
\end{array}\right ]\label{matrix}
\end{eqnarray}
where all matrix elements are i.i.d. complex random variables. Here we
use the uniform distribution in the unitary disk of the complex plane. 
This implies 
that the eigenvalue density of the ensemble is only a function 
of the modulus of the eigenvalue. The eigenvalues of a sample matrix of
size $n=800$ are shown in Fig.\ref{a1} (left).

We next consider two deformations of the matrix $M$, by a complex 
parameter $z=e^{\xi+i\varphi}$: 
\begin{eqnarray}
M(z^n) =\left [\begin{array}{cccc}
a_1 &    b_1   &   {}  & z^nc_1\\
c_2 & \ddots &\ddots   & {}    \\
{}  & \ddots & \ddots  & b_{n-1}   \\
b_n/z^n&    {}  &   c_n   & a_n   \\
\end{array}\right ], \\
M_b(z) =\left [\begin{array}{cccc}
a_1 &    b_1/z   &   {}  & zc_1\\
zc_2 & \ddots &\ddots   & {}    \\
{}  & \ddots & \ddots  & b_{n-1}/z   \\
b_n/z&    {}  &   zc_n   & a_n   \\
\end{array}\right ].\label{matrices}
\end{eqnarray}
The two matrices are similar, $M_b(z)=SM(z^n)S^{-1}$, through a diagonal 
matrix with entries $S_{ii}=z^i$. The balanced matrix $M_b(z)$ is more 
convenient for numerical work. Since the matrices share the same
set of eigenvalues, a rotation of $z$ by $2\pi/n$ does not change the 
eigenvalues of $M_b(z)$. 

The eigenvalues of $M_b(z)$ are shown in Fig.\ref{a1} (right). 
The distribution looks remarkable: a ``circle'' centered in the origin 
bounds an outer annular halo where the eigenvalues appear exactly 
in the same positions as those in the left figure. The inner region is 
void: all the eigenvalues that were there before deformation ($\xi=0$) have 
moved to the boundary circle. 
\begin{figure}
\begin{center}
\includegraphics[width=6.5cm]{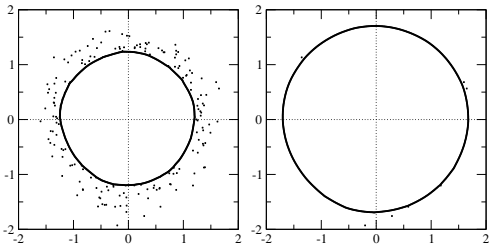}
\caption{\label{a2} The same random matrix entries as in Fig.\ref{a1}
with $\xi =0.7$ and $\varphi=0$ (left), $\xi=1$ and $\varphi =0$ (right).}
\end{center}
\end{figure}
As $|z|$ becomes larger, see Fig.\ref{a2}, the circle enlarges as well,
but the eigenvalues in the annular halo do not move, until they are 
swept by the circle. For large $\xi $ only the circle remains.
This is not surprising: in the 
limit of large $|z|$ the matrix $M_b(z)$ simplifies to bidiagonal. The 
eigenvalue equation can be solved explicitly and gives 
$E^n= z^n b_1\cdots b_n$. Then the eigenvalues 
$E_k = |z|e^{\langle \ln |b|\rangle } \exp i(\theta+2\pi k/n)$ are equally 
spaced and lie on a circle of radius $r$ such that 
$\log r=\xi+\langle \log |b|\rangle $.

\begin{figure}
\begin{center}
\includegraphics[width=3.5cm]{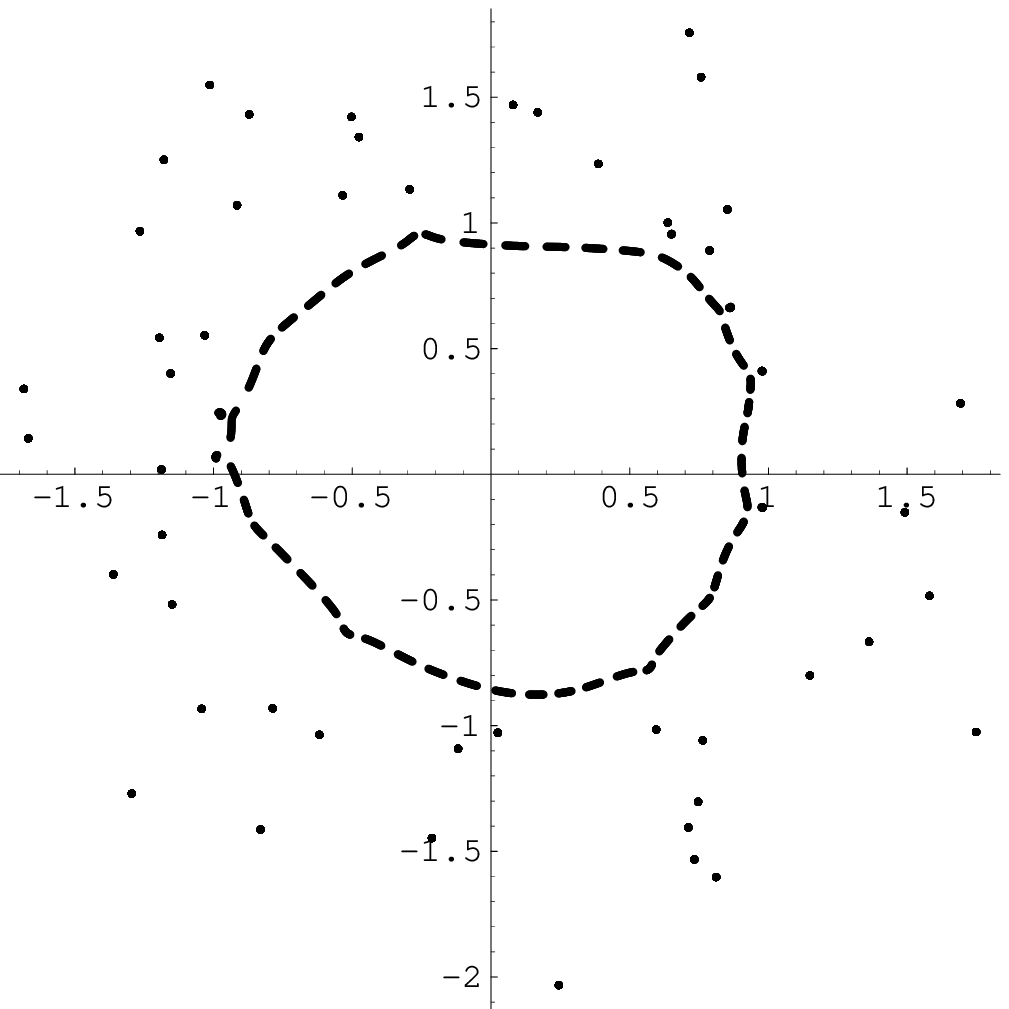}
\hspace{0.5cm}
\includegraphics[width=3.5cm]{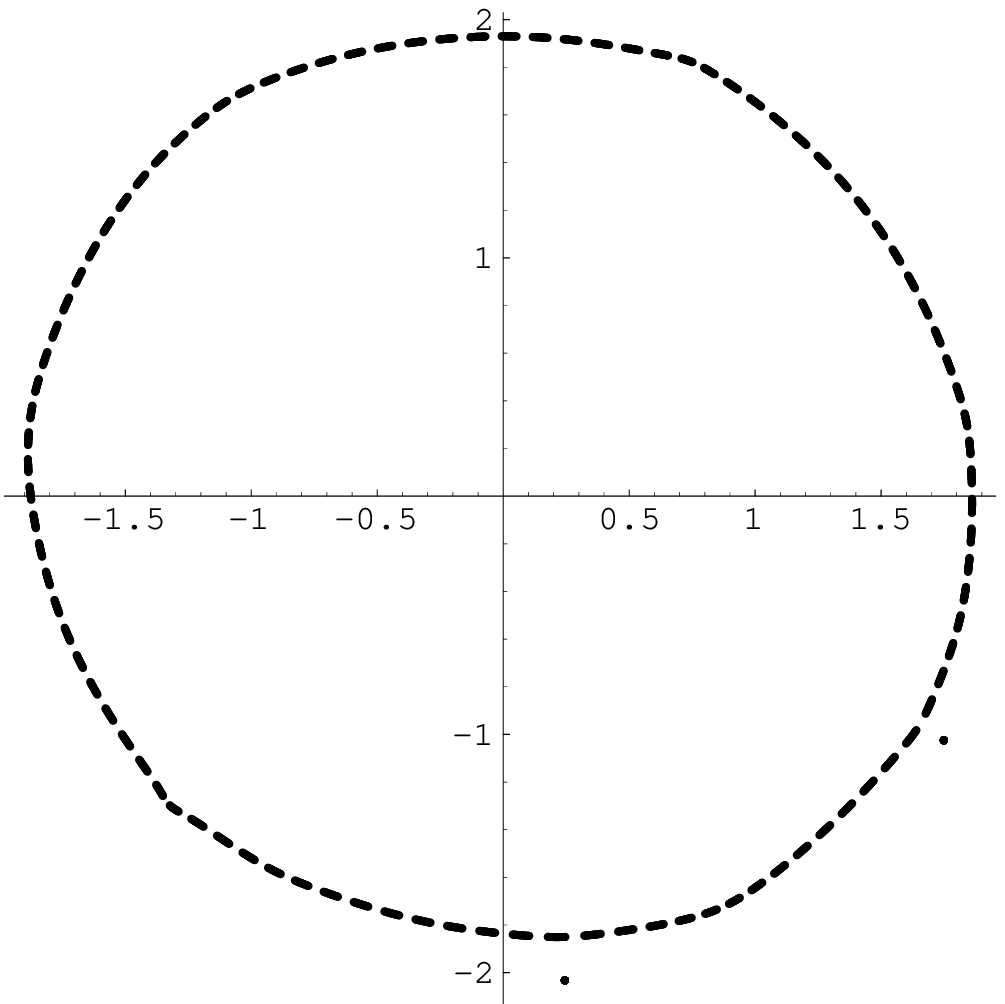}
\caption{\label{a3} 
Motion in the complex plane of the eigenvalues of a single non Hermitian
tridiagonal matrix of size $n=100$ as the phase $\varphi $ is changed in 
half of its range. The same matrix entries are used,  with $\xi=0.5$ (left) 
and $\xi=1$ (right). 
Eigenvalues not belonging to the loop are seen to be fixed, corresponding to 
localized states. The loop becomes closed if the whole angular range is 
evaluated.}
\end{center}
\end{figure}

Eigenvalues on the circle and in the halo respond differently to
the phase $\varphi = {\rm arg}\,z$. As $\varphi $ sweeps the Brillouin 
zone from 0 to $2\pi/n $, only the eigenvalues sitting on the circle move 
(and remain therein), while the outer ones do not have measurable changes 
at all. This is illustrated in Fig.\ref{a3}, which also shows that an
eigenvalue on the circle moves to the position of a neighboring one 
as $\varphi$ is increased by $2\pi/n$. 
\begin{figure}
\begin{center}
\includegraphics[width=3cm]{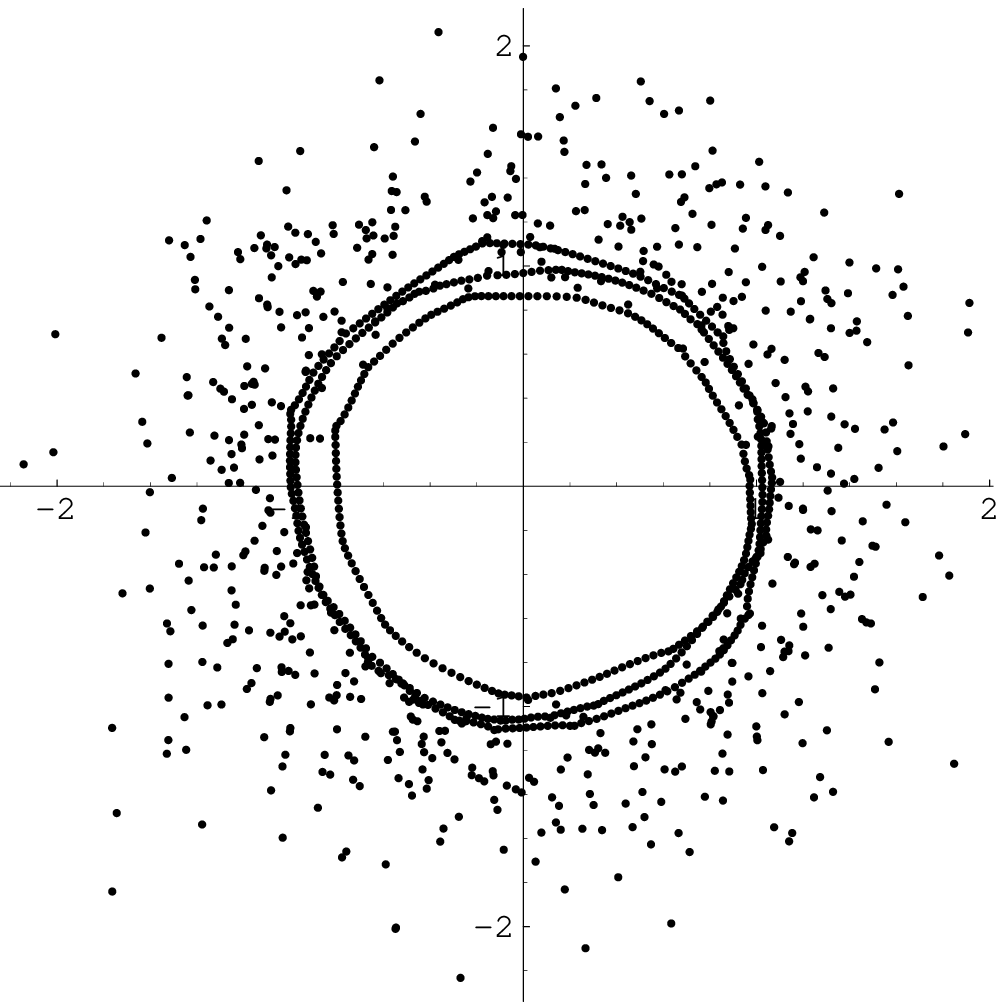}
\hspace{0.5cm}
\includegraphics[width=3cm]{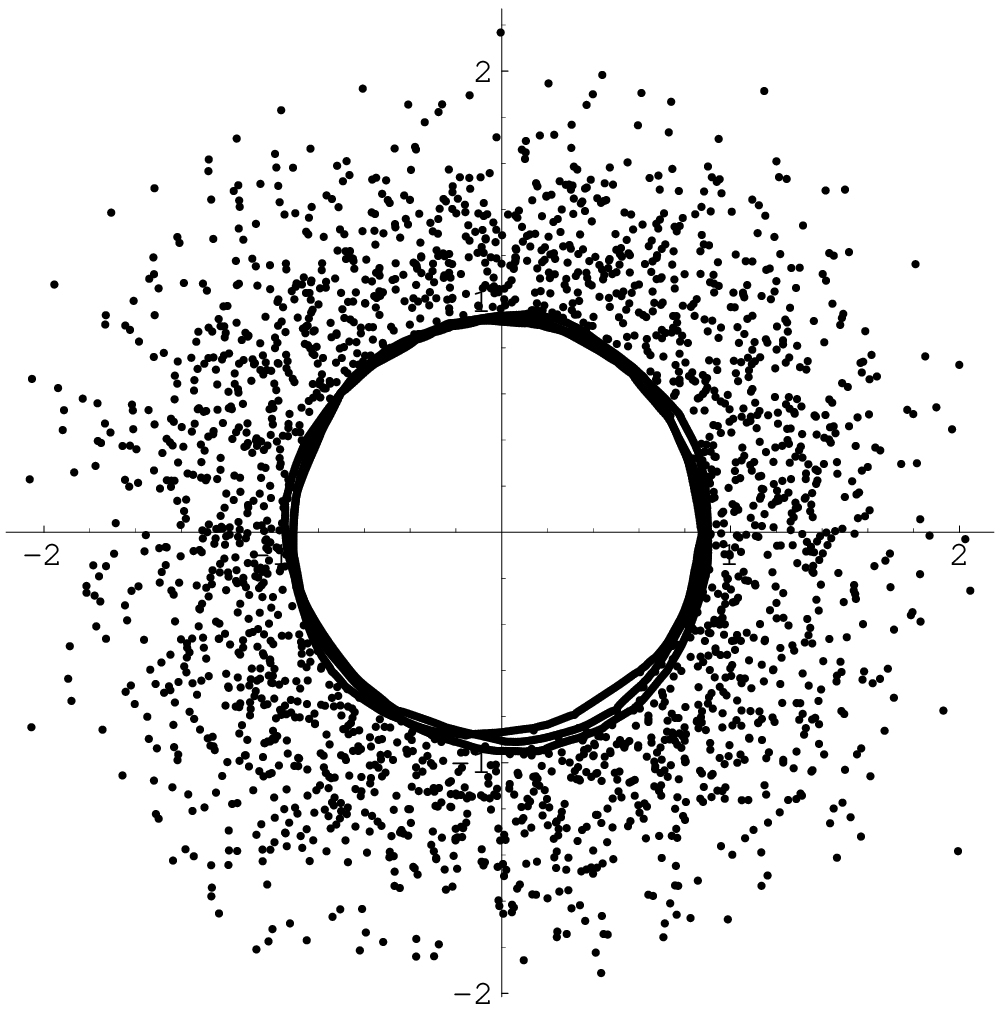}
\caption{\label{a111} 
Eigenvalues in the complex plane of 3 different samples of size $n=400$ 
(left)
and $n=1200$ (right). All matrices with $\xi =0.5$ and $\varphi=0$. }
\end{center}
\end{figure}
By increasing the size $n$ of the matrices, the ``circle'' is seen to become 
independent of the sample and more regular, (Fig.\ref{a111}).
\begin{figure}
\begin{center}
\includegraphics[width=4cm]{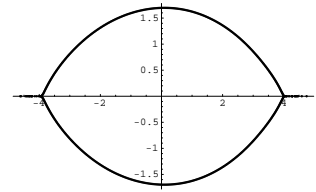}
\caption{\label{HNfig} 
Eigenvalues of a single Hatano-Nelson tridiagonal matrix of 
size $n=600$, with $\xi =1$, $b_k=c_k=1$ and random numbers $a_k$ 
uniformly distributed in $(-3.5,3.5)$. They belong to two arcs or
two wings in the real axis, which are the residuate of the undeformed 
spectrum.}
\end{center}
\end{figure}

The phenomenon described is analogous to what Hatano and 
Nelson\cite{Hatano96} discovered for random tridiagonal Hermitian matrices 
($a_k$ real, $c_{k+1}=b_k^*$) where the undeformed eigenvalues 
($\xi=0$) are real. The deformation forces them to move into the complex 
plane and distribute along a two-arc loop, with possible external wings of 
unaffected eigenvalues in the real axis (Fig.\ref{HNfig}). 
The two-arc loop and wings of the Hermitian model correspond to the 
circle and annular halo of the non-Hermitian model discussed here.

In Section II and III we study the spectral density of the undeformed ensemble
and the localization of eigenvectors, measured by the Lyapunov exponent
or by the variance. In Section IV we explain the observed 
spectral features of the deformed ensemble by means of the Argument 
Principle of complex analysis and a spectral duality between
the eigenvalues of $M(z^n)$ and those of the transfer matrix.

\vskip1truecm
\begin{figure}
\begin{center}
\includegraphics[width=7cm]{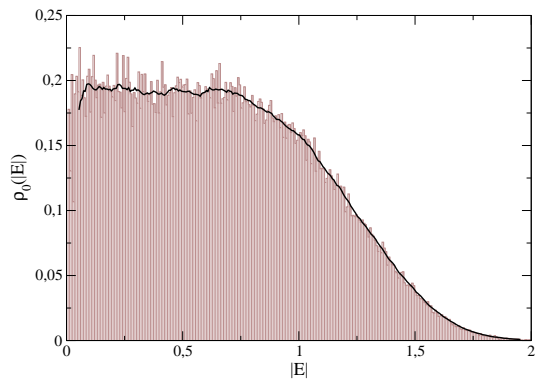}
\caption{\label{hystog} 
Eigenvalue density $\rho_0(|E|)$ of the undeformed ensemble (matrix size 
$n=1000$, 100 samples). The dark line is obtained by smoothing the hystogram 
data on local windows of 10 bins (out of 300). The plateau is fitted 
by the value $0.193(2)$.}
\end{center}
\end{figure}

\section{The spectrum of $M$}
Since the matrix entries of $M$ are chosen to be uniformly distributed in 
the unit complex disk, the average eigenvalue density of the ensemble, 
$\rho_0(E)=\langle \frac{1}{n}\sum_i\delta_2 (E-E_i)\rangle $,
depends on $|E|$. 
The eigenvalue equation
\begin{equation}
c_k\, u_{k-1} + a_k\, u_k + b_k\, u_{k+1} = E\, u_k \label{eigenvalue_equation}
\end{equation}
written for the component $u_k$ with highest absolute value, implies 
the inequality $|E|\le |a_k|+|b_k|+|c_k|\le 3$. Thus the disk that supports 
the density has a radius not exceeding $3$; the numerical evidence 
is that it has length $2$.

We diagonalized 100 matrices of size $n=1000$ to obtain 
numerically the density of eigenvalues $\rho_0(|E|)$ shown in Fig.\ref{hystog}.
The lowest moments $\mu_k=\langle \frac{1}{n}\sum_i|E_i|^k\rangle $=
$2\pi\int_0^2 dx x^{k+1}\rho_0(x)$ 
are also evaluated: $\mu_1=0.9107(12)$, $\mu_2=0.9678(22)$, $\mu_3=1.1327(36)$
and $\mu_4=1.4204(59)$.
\begin{figure}
\begin{center}
\includegraphics[width=7cm]{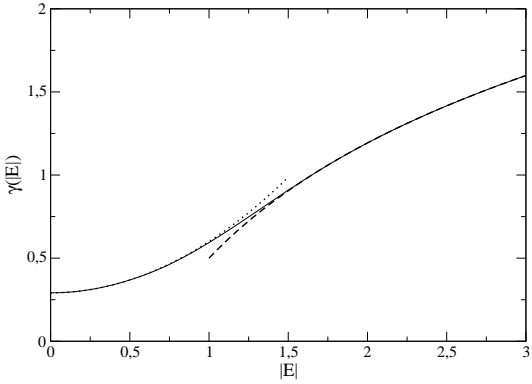}
\caption{\label{gamma} 
The Lyapunov exponent $\gamma (|E|)$ evaluated numerically (100 matrices of 
size $n=1000$). The two curves are a quadratic fit near the origin,
$0.2914(1)+0.309(0)|E|^2$, and the exact analytic expression $\lg (|E|)+1/2$
of $\gamma $ for $|E|>2$. The fit near the origin 
is consistent with the value found for $\rho_0(0)$ (via Poisson equation).}
\end{center}
\end{figure}

\begin{figure}
\begin{center}
\includegraphics[width=7cm]{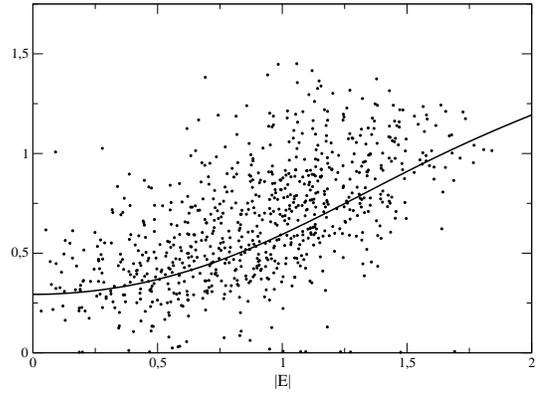}
\caption{\label{gamma-var} 
Pairs $(|E_a|,\gamma_a)$, where $E_a$ are eigenvalues and $\gamma_a$
are rates of exponential localization, for the eigenvectors of a matrix $M$
of size $n=800$ ($\xi =0$). The continuous line is the Lyapunov exponent
$\gamma (|E|)$.}
\end{center}
\end{figure}

\section{The Lyapunov exponent}
The numerical evaluation of the eigenvectors of $M$ shows that they are 
strongly localized for all eigenvalues; indeed, for large matrix size, 
they decay exponentially (Anderson localization). The rate of decay is 
measured by the Lyapunov exponent, an asymptotic property of transfer 
matrices. 

The transfer matrix of a realization $M$ is the product of $2\times 2$ 
random matrices:
\begin{eqnarray}
t(E) = \prod_{k=1}^n \left [\begin{array}{cc} 
b_k^{-1}(E -a_k) & -b_k^{-1}c_k\\ 1 & 0\\ \end{array}\right ].
\end{eqnarray}
Its eigenvalues can be written as $z_{\pm}^n=e^{n(\xi_\pm +i\varphi_\pm)}$.  
For large $n$ the exponents $\xi_\pm (E)$ become opposite: 
$\xi_+ + \xi_- $ $= \frac{1}{n}\log |\det t(E)| $ $= 
\frac{1}{n}\sum_{k=1}^n \log|c_k|-\log |b_k|\to 0$.
 
According to the theory of random matrix products, for large $n$ the positive
exponent $\xi_+ $ becomes independent of $n$ and the realization of randomness,
and converges to the {\em Lyapunov exponent} of the matrix ensemble. 
The Lyapunov exponent can be evaluated by an extension of Thouless formula to 
non-Hermitian matrices\cite{Derrida00,Goldsheid05},
\begin{eqnarray}
\gamma (E)= \int d^2 E' \rho_0 (E')\log |E-E'| -\langle\log |b|\rangle,
\end{eqnarray}
where $\rho_0$ is the eigenvalue density of the ensemble of matrices $M$.
Note that for complex spectra, the equation for $\gamma $ implies the Poisson 
equation $\nabla^2 \gamma (E) =2\pi \rho_0(E)$.
Therefore $\gamma (E)$ can be understood as the electrostatic potential 
generated by a charge distribution in the plane with density $\rho_0 (E)$.  

For a distribution of matrix entries that is uniform in the unit disk, 
it is $\langle \log |b|\rangle =-1/2$ and $\rho_0$ is rotation invariant. 
Then the integral can be simplified:
\begin{eqnarray}
\gamma (|E|) =&& \log |E|\, {\cal N}_0(|E|)\nonumber \\ 
&&+ 2\pi \int_{|E|}^\infty dE'E'\rho_0(E')\log E' +1/2. \label{thouless}
\end{eqnarray}
The integral $\int_0^{2\pi}\log|r-r'e^{i\varphi}|d\varphi =2\pi \log
{\rm max}(r,r')$ was used. ${\cal N}_0(|E|)$ is the fraction of the spectrum 
inside the disk of radius $|E|$. For $|E|$ larger than the spectral 
radius it is 
\begin{eqnarray}
\gamma(|E|) = \log (|E|) + 1/2, \quad |E|\ge 2.
\end{eqnarray}

The Lyapunov exponent is an increasing function of $|E|$. Its
numerical evaluation is shown in Fig.\ref{gamma}.

We checked numerically the exponential decay of eigenvectors with a rate
given by $\gamma (|E|)$. 
If $\vec u$ is an eigenvector of $M$, with components $\{ u_k\}_{k=1}^n$,
the numbers $|u_k|^2$ provide the probability distribution for the position 
of a particle in the lattice $1\ldots n$. 
We choose to measure the localization 
of the eigenvector by the variance in position:
\begin{equation}
{\rm var} [\vec u]=
\left ( \sum_{k=1}^n |u_k|^2 (k-\overline k)^2 \right)^{1/2}, \label{var}
\end{equation} 
where $\overline k = \sum_k k|u_k|^2$ is the mean position of the particle.
Other measures could be used, as the inverse participation ratio or the
Shannon entropy; in this case the eigenvectors are peaked on small intervals, 
and the variance has a clear meaning.

For an ideal state $\vec v$ that is exponentially localized, 
$|v_k|^2 =(\tanh \gamma) \,e^{-2\gamma |k|} $, ($\gamma n\gg 1$), 
the variance is ${\rm var} [\vec v] = \sinh (1/\gamma)$. We use the same
relation to compute a rate $\gamma_a$ from the numerically evaluated
variance of an eigenstate $\vec u_a$.
In Fig.\ref{gamma-var} we plot the numerical pairs $(|E_a|,\gamma_a)$ for 
the eigenvalues and eigenvectors of a single matrix $M$ of size $n=800$,
together with the Lyapunov exponent $\gamma(|E|)$, given by Thouless formula
(\ref{thouless}).
The numerical data are consistent with the picture of exponential 
localization of eigenvectors.

\begin{figure}
\begin{center}
\includegraphics[width=7cm]{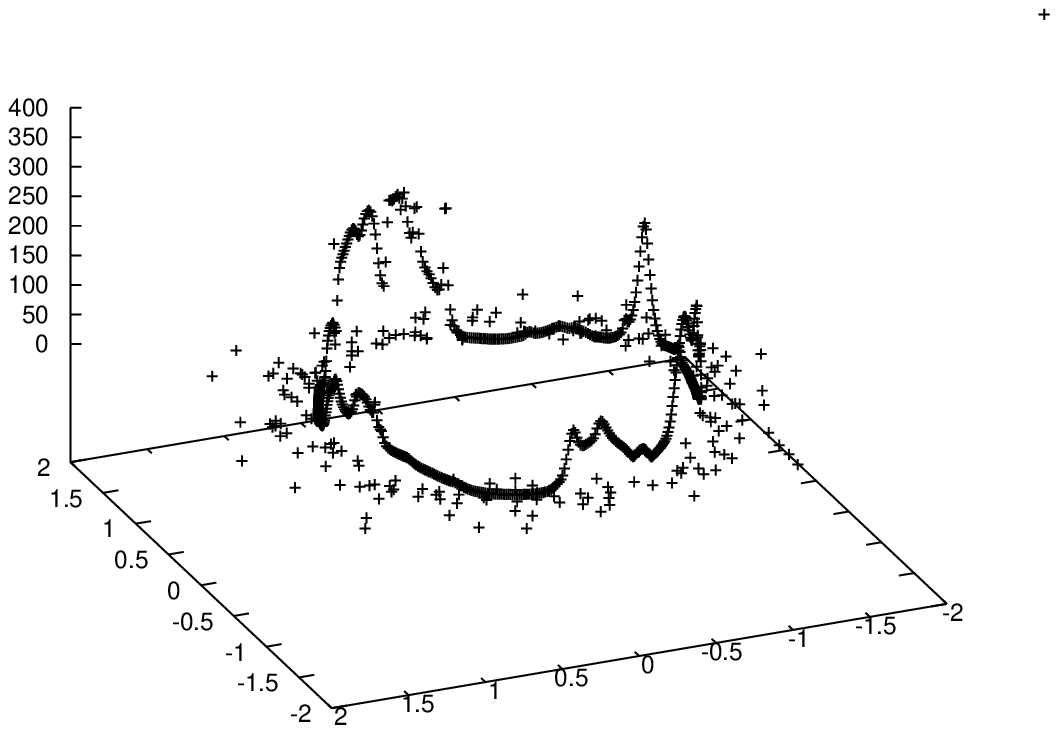}
\caption{\label{3D}Variance of eigenvectors ($z$-axis) and 
corresponding eigenvalues ($x=Re E$, $y=Im E$) of a random matrix $M_b(e^\xi)$ 
of size $n=800$ and $\xi =\log 2$. The radius of the hole is approximately
1.16.}
\end{center}
\end{figure}

\section{Hole, halo and localization}
When the parameter $\xi $ is switched on, it appears in the corners
of the matrix $M(z^n)$ and modifies the boundary conditions in 
(\ref{eigenvalue_equation}). In the transition from
$M$ to $M(z^n)$, one expects that the eigenvalues of enough localized 
eigenstates do not change. An eigenstate $\vec u=\{u_k\}$ of $M(z^n)$ 
corresponds to an eigenstate $S\vec u$ of $M_b(z)$, with components 
$z^k u_k$. If $|u_k|\approx e^{-\gamma |k|}$ for large $k$, the factor
$e^{k\xi}$ delocalizes it if $\gamma <\xi$. This simple argument by 
Hatano and Nelson indicates a threshold value $\gamma (|E|)=\xi$ at 
which eigenvalues must be drastically influenced by the deformation.

In Fig.\ref{3D} we plot the variances ($z$ axis) of the eigenvectors of a 
matrix $M_b(z=2)$ of size $n=800$, and the corresponding complex 
eigenvalues (horizontal plane). The boundary of the circular hole is 
populated by the eigenvectors which are delocalized.

The existence of an empty disk and a halo of fixed eigenvalues
for the deformed ensemble thus reflects the localization properties of the 
eigenvectors of $M$ as a function of $|E|$, i.e. the function $\gamma(|E|)$. 

{\bf Proposition}: in the large $n$ limit, $M_b(e^{\xi+i\varphi})$ has no
eigenvalues in the disk of radius $r$, where
\begin{equation}
\gamma (r) = \xi
\end{equation}
Proof: The hole in the spectrum of $M_b(z)$ can be understood via the 
{\em Argument Principle} of complex analysis: the number of zeros of the 
analytic function $f(E)=\det[E-M_b(z)]$ inside a disk of radius $r$ is
equal to the variation of arg$f(E)/2\pi$ along the contour of the 
disk.

The function $f(E)$ is related to the eigenvalues $z^n_\pm (E)$ of the 
transfer matrix $t(E)$ by a duality identity\cite{Molinari08,Molinari}:
\begin{eqnarray}
\det[E-M_b(z)] = -\frac{1}{z^n}(b_1\cdots b_n)\det[t(E)-z^n]\label{dual}
\end{eqnarray}
Then
\begin{eqnarray}
{\rm arg}\det [E-M_b(e^{\xi+i\varphi})]= {\rm const.}\nonumber \\
+ {\rm arg}[e^{n(\xi_+-\xi)+in(\varphi_+ -\varphi)}-1]\nonumber\\
+  {\rm arg}[e^{n(\xi_--\xi)+i(\varphi_- -\varphi)}-1]\nonumber
\end{eqnarray}
Let us fix $\xi>0$ and take the large $n$ limit. Then: $\xi_+>0$ and 
$\xi_-<0$;
$ {\rm arg}[e^{n(\xi_+-\xi)+in(\varphi_+-\varphi)}-1]$ 
equals $n(\varphi_+-\varphi)$ if $\xi_+>\xi$, and $\pi$ if $\xi>\xi_+$; 
$ {\rm arg}[e^{n(\xi_--\xi)+in(\varphi_-\varphi)}-1]=\pi $ always. 
We also identify $\xi_+(E)$ with the Lyapunov exponent $\gamma(|E|)$. 
The variation of arg$f(E)/2\pi$ along a circumference of radius 
$r$ is zero if $\xi>\gamma(r)$. $\Box$

Since $\gamma(0)$ is nonzero, there is a threshold value 
$\xi_{min}\approx 0.291$ below which no hole opens in the spectral
support.  

\section{Conclusions}
The Hatano Nelson deformation opens a hole in the spectrum of non Hermitian
tridiagonal random matrices with i.i.d. matrix elements. The eigenvalues
that are swept to the boundary of the hole 
correspond to states that are no longer Anderson localized.
This is explained in terms of a spectral duality, stability of the Lyapunov
exponent, and the Argument Principle.

Tridiagonal matrices with different strengths of randomness in the three
diagonals would also show similar spectral features.

\vskip0.5truecm
{\bf Aknowledgements}: L.G.M. wishes to thank prof. I. Goldsheid for 
having inspired him the present investigation, and for useful comments.


\vfill
\end{document}